\documentclass[twocolumn]{aastex631}

\usepackage{CJKutf8}

\usepackage[caption=false]{subfig}
\usepackage{tabularx}
\usepackage{capt-of}
\usepackage{hhline}
\usepackage{amsmath}	
\usepackage{amssymb}	
\usepackage{textcomp}
\usepackage{upgreek}

\newcommand{\cmark}{\ding{51}} 
\newcommand{\xmark}{\ding{55}} 
\usepackage{pifont} 

\newcommand{\Msun}{\mathrm{M}_{\odot}}
\newcommand{\fPBH}{f_{\rm PBH}}

\begin{document}

\title{How do Primordial Black Holes change the Halo Mass Function and Structure?}

\author[0000-0003-1541-177X]{Saiyang Zhang\begin{CJK*}{UTF8}{bsmi}(張賽暘）\end{CJK*}}

\affiliation{Department of Physics, University of Texas at Austin, Austin, TX 78712, USA}

\affiliation{Weinberg Institute for Theoretical Physics, Texas Center for Cosmology and Astroparticle Physics, \\ University of Texas at Austin, Austin, TX 78712, USA}
\author[0000-0003-0212-2979]{Volker Bromm}
\affiliation{Department of Astronomy, University of Texas at Austin, Austin, TX 78712, USA}
\affiliation{Weinberg Institute for Theoretical Physics, Texas Center for Cosmology and Astroparticle Physics, \\ University of Texas at Austin, Austin, TX 78712, USA}

\author[0000-0002-4966-7450]{Boyuan Liu\begin{CJK*}{UTF8}{bsmi}(劉博遠)\end{CJK*}}
\affiliation{Department of Astronomy, University of Texas at Austin, Austin, TX 78712, USA}
\affiliation{Institute of Astronomy, University of Cambridge, Madingley Road, Cambridge, CB3 0HA, UK}
\affiliation{Universit\"at Heidelberg, Zentrum fur Astronomie, Institut f\"ur Theoretische Astrophysik, D-69120 Heidelberg, Germany}



\begin{abstract}
We examine the effects of massive primordial black holes (PBHs) on cosmic structure formation, employing both a semi-analytical approach and cosmological simulations. Our simulations incorporate PBHs with a monochromatic mass distribution centered around $10^6 \ \rm M_{\odot}$, constituting a fraction of $10^{-2}$ to $10^{-4}$ of the dark matter (DM) in the universe, with the remainder being collision-less particle dark matter (PDM). Additionally, we conduct a $\Lambda$CDM simulation for comparative analysis with runs that include PBHs. At smaller scales, halos containing PBHs exhibit similar density and velocity dispersion profiles to those without PBHs. Conversely, at larger scales, PBHs can expedite the formation of massive halos and reside at their centers due to the `seed effect'. To analyze the relative distribution of PBH host halos compared to non-PBH halos, we apply nearest-neighbor (NN) statistics. Our results suggest that PBH host halos, through gravitational influence, significantly impact the structure formation process, compared to the $\Lambda$CDM case, by attracting and engulfing nearby newly-formed minihalos. Should PBHs constitute a fraction of DM significantly larger than $\sim$$10^{-3}$, almost all newly-formed halos will be absorbed by PBH-seeded halos. Consequently, our simulations predict a bimodal feature in the halo mass function, with most of the massive halos containing at least one PBH at their core and the rest being less massive non-PBH halos.

\end{abstract}



\section{Introduction} \label{sec:intro}

Since the James Webb Space Telescope (JWST) deployment in 2022, groundbreaking observations have begun to illuminate the enigmatic early universe and the genesis of its first galaxies. Recent JWST observations have revealed an astonishing population of massive galaxies with stellar masses exceeding $10^9\,\Msun$ at redshifts $z\gtrsim 10$ \citep[e.g.,][]{Maisies:2022,GHz2,GLASSz13,Yan2023ApJ...942L...9Y, Castellano2023ApJ...948L..14C, Labbe2023Natur.616..266L, Yan2023arXiv231115121Y}{}{}, and supermassive black holes (SMBHs) with masses $\gtrsim10^7\,\Msun$ at $z = 7-10$ \citep[e.g.,][]{Goulding2023ApJ...955L..24G,Bogdan:2023UHZ1, Larson_2023_BH,Maiolino2023,Greene2024,Natarajan:2023UHZ1, Kovacs2024ApJ...965L..21K}{}{}. While most galaxies are identified through photometric measurements subject to calibration uncertainties, a select few have been spectroscopically confirmed \citep[i.e.][]{Curtis-Lake2023NatAs...7..622C, Arrabal2023ApJ...951L..22A, Robertson2023NatAs...7..611R}{}{}. The existence of those galaxies, with higher than expected stellar mass at such high redshift (e.g., GN-z11; \citealt{Tacchella_2023}), challenges our standard $\Lambda$CDM cosmology \citep{Inoyashi2022ApJ...938L..10I, Boylan2023}. To alleviate this tension, a re-calibration of the stellar mass has been proposed, such as using different stellar initial mass functions and their resulting spectral energy distributions, when interpreting the observations \citep[e.g.,][]{Finkelstein2023,Wang2024ApJ...963...74W}. 

A related challenge is posed by the discovery and follow-up spectroscopic confirmation of massive black holes in the pre-reionization universe (e.g., UHZ1; \citealt{Goulding2023ApJ...955L..24G, Bogdan:2023UHZ1}). The standard SMBH formation channel is via gas accretion onto the light black hole seeds from massive Population~III (Pop~III) star remnants with $\sim 100-1000\ \Msun$ ~\citep[for reviews, see][]{Smith2019:BHreview, Inayoshi:2020}{}{}. However, rapid growth of the light seeds to $10^7\,\Msun$ in less than $\sim 500$\,Myr by $z \sim 10$ requires an accretion rate exceeding the Eddington limit \citep[e.g.,][]{Bogdan:2023UHZ1}{}{}. 
Recent observations of high-redshift active galactic nuclei (AGN) suggest that the majority may be dormant with sub-Eddington accretion rates \citep[e.g.,][]{Jeon2023}. If interspersed with episodic super-Eddington phases, however, such light-seed models may be able to explain at least some of the newly discovered high-$z$ SMBHs \citep{Juodzbalis2024arXiv240303872J,Maiolino2024Natur.627...59M}. 
An alternative SMBH formation channel proceeds via heavy black hole seeds from the direct collapse of a primordial gas cloud \citep{DCBH2003ApJ...596...34B, Begelman2006, Lodato2006}. The resulting accelerated growth \citep[e.g.,][]{Jeon2024arXiv240218773J} may be required to explain part of the emerging JWST phenomenology \citep{Bogdan:2023UHZ1, Kovacs2024ApJ...965L..21K}.

To understand the challenges posed by the JWST observations for early structure formation, we need to consider the properties of the dominant dark matter (DM) component. Of particular interest are models where a fraction of dark matter is composed of massive primordial black holes (PBHs), with the possibility of accelerating the early formation of massive galaxies \citep[e.g.,][]{Boyuan2022ApJ,Boyuan2023arXiv231204085L,Su2023,Colazo2024}, or of providing the initial seeds for high-$z$ SMBHs \citep[e.g.,][]{Silk2024ApJ...961L..39S}.

PBHs are well motivated DM candidates, theorized to form in the early universe through various mechanisms \citep[reviewed in][]{Carr2020ARNPS..70..355C}. These include the collapse of overdensities prior to matter-radiation equality \citep{Zeldovich1967SvA....10..602Z, hawking1971gravitationally, Escriva2022} and collapse of domain walls during inflation \citep{Belotsky2019}. While PBHs may not account for all dark matter, even a sub-dominant mass fraction can significantly influence cosmic history and imprint observable signals \citep[for a review of PBH phenomenological constraints, see][]{Carr2021}{}{}. Especially, in structure formation history, PBHs play an important role through the `Poisson' and `seed' effects \citep{Carr2018MNRAS.478.3756C, Inman2019PhRvD.100h3528I}. The 'Poisson' effect stems from the number density fluctuations of discrete PBHs, which can accelerate the growth of overdensities and the formation of halos \citep{Meszaros1975A&A....38....5M, Afshordi2003ApJ...594L..71A, Kashlinsky2021PhRvL.126a1101K,Cappelluti2022ApJ}.  On smaller scales, the `seed' effect arises from individual PBHs, where they act as seeds for SMBH and galaxy formation by 
accretion of the surrounding material \citep{Mack2007ApJ, Ricotti2007ApJI, Ricotti2008ApJII}. 

Previously, we have studied the influence of stellar-mass PBHs ($10-100\,\Msun$) on the formation of Pop~III stars \citep{Boyuan2022MNRAS.514.2376L}, and the contribution of PBH DM to cosmic radiation backgrounds (\citealt{Zhang2024MNRAS.528..180Z}; see also \citealt{Ziparo2022}). We have also considered the extreme case of supermassive ($\sim 10^9-10^{11}\,\Msun $) PBHs on the assembly of the first galaxies \citep{Boyuan2022ApJ,Boyuan2023arXiv231204085L}. All of these studies indicate that structure formation is indeed accelerated in PBH cosmologies, with an earlier formation of DM halos. In this paper, motivated by the recent JWST observational results, we explore the influence of $10^6\,\Msun$ PBHs on structure formation, via a combination of `Poisson' and `seed' effects. This PBH mass scale is motivated by a formation scenario that invokes the softening of the equation of state during $e^+e^-$ annihilation, when the universe had cooled to a temperature of $\sim 1\rm \ MeV$ \citep{Carr2021PDU....3100755C}.

We run our simulations till $z\simeq 10$, where our results can be compared with JWST observations, exploring PBH mass fractions of $10^{-2}$ to $10^{-4}$, and assuming that the remainder consists of particle dark matter (PDM). When evaluating the resulting halo mass functions, we interpret any deviations from the Press-Schechter (PS) formalism \citep{Press1974ApJ} through a nearest-neighbor (NN) statistical approach \citep{Banerjee2021MNRAS.500.5479B}, and develop a simple heuristic model to understand the structural changes induced by different PBH cosmologies.

In Section~\ref{sec:hmf}, we outline our analytical framework designed to assess the influence of PBHs on structure formation and the halo mass function. Following this, Section~\ref{sec:sim} details the initial settings and relevant parameters of our simulations. Based on these simulations, in Section~\ref{sec:result}, we derive the DM halo abundances within the different models, and explore the  impact of PBHs across various scales. We conclude in Section \ref{sec:conclusion}, where we summarize our findings and discuss directions for future research.

Throughout the paper, we employ the cosmological parameters from the \textit{Planck15} measurement \citep{Plank2016A&A...594A..13P}, as follows: $\Omega_{\rm m} = 0.3089$, $h = 0.6774$, $\sigma_8 = 0.8159$, $n_s = 0.9667$.

\section{PBH Halo Mass Function}\label{sec:hmf}
For simplicity, we assume a monochromatic mass distribution for PBHs with $M_{\rm PBH} = 10^6\,\Msun$. Observations constrain the abundance of PBHs as a fraction of total DM mass, denoted by $\fPBH$, leading to maximum values allowed for a monochromatic distribution \citep{Carr2021}. The current upper limit on $10^6\,\Msun$ PBHs is $\fPBH \lesssim 10^{-3}$, deriving from the effect of dynamical friction \citep{Carr1999ApJ...516..195C} and the formation of large-scale structure \citep{Carr2018MNRAS.478.3756C}. We also note that the most stringent limit arises from $\mu$-distortion of the cosmic microwave background (CMB) when assuming PBH formation from the high-density tail of a Gaussian random field \citep{Chluba2012ApJ...758...76C,Nakama2018PhRvD..97d3525N,Chluba2021ExA....51.1515C,Hooper2024}, and the future mission PIXIE will further refine the measurement of the CMB power spectrum on small scales \citep{PIXIE2024arXiv240520403K}. On the other hand, relaxing the Gaussian fluctuation assumption \citep{Nakama2018PhRvD..97d3525N} and considering formation mechanisms dependent on a different inflation field  \citep[i.e.][]{Kawasaki2019PhRvD.100j3521K, Carr2020ARNPS..70..355C}{}{} could alleviate this bound. Therefore, in this paper, we adopt $10^{-3}$ as the fiducial value, consistent with existing bounds, and vary the PBH fraction in DM within $\fPBH \sim 10^{-2} - 10^{-4}$ to study its impact on the halo mass function (HMF). 

Following previous work \citep{Inman2019PhRvD.100h3528I,Boyuan2022MNRAS.514.2376L, Boyuan2023arXiv231204085L, Zhang2024MNRAS.528..180Z}, we briefly summarize the analytical formalism for deriving the HMF within PBH cosmologies, and we reproduce the key equations for the convenience of the reader.

The average (comoving) PBH number density is 
\begin{displaymath}
\bar{n}_{\mathrm{PBH}}  =\frac{f_{\mathrm{PBH}} }{M_{\rm PBH}} (\Omega_{\mathrm{m}} - \Omega_{\mathrm{b}}) \frac{3 H_0^2}{8 \pi G}\mbox{\ ,}
\end{displaymath}
defining $k_{\mathrm{PBH}}=\left(2 \pi^2 \bar{n}_{\mathrm{PBH}}\right)^{1 / 3}$ as the inter-PBH separation scale, and $\Omega_{\rm b}$ is the energy density fraction of baryons. PBHs are assumed to be randomly distributed over space, and the fluctuation in PBH number density leads to the `Poisson' effect, causing isocurvature fluctuations in the density field \citep{Carr2018MNRAS.478.3756C}. Previous treatments by \cite{Afshordi2003ApJ...594L..71A, Kashlinsky2016ApJ...823L..25K} assumed that PBHs provide all the DM. Here, we relax this assumption by considering $f_{\mathrm{PBH}}$ as a free parameter. We express the total power spectrum $P_{\mathrm{tot}}(k)$, extrapolated to $z=0$, as the combination of adiabatic $P_{\rm ad}(k)$ and isocurvature $P_{\rm iso}(k)$ modes with a mode mixing term $P_{\rm corr}(k)$ \citep{Inman2019PhRvD.100h3528I, Boyuan2023arXiv231204085L}, as
\begin{equation}
 \begin{aligned}
    P_{\mathrm{tot}}(k)&=P_{\rm ad}(k) + P_{\rm iso}(k) + P_{\rm corr}(k)\,,
 \end{aligned}
\label{eq:pk}
\end{equation}
 where $P_{\rm ad}(k)$ refers to the standard $\Lambda$CDM power spectrum, generated with the \textsc{python} package \href{https://bdiemer.bitbucket.io/colossus/index.html}{\textsc{colossus}} \citep{Diemer2018ApJCOLOSSUS}. The isocurvature term $P_{\rm iso}(k)$ arises from the Fourier transform of the overdensities attributable to the discrete nature of PBHs:
 \begin{equation}
         P_{\rm iso}(k) = \left[f_{\rm PBH}D_{0}\right]^{2}/\bar{n}_{\rm PBH},
         \label{eq:piso}
 \end{equation}
and it becomes significant at scales smaller than the matter-radiation equality scale $k_{\rm eq} \sim 0.01\,\mathrm{Mpc}^{-1}$. Here $D_0 = D_{\rm PBH} (a = 1)$ is the growth factor of PBH-induced perturbations at $z = 0 $, well approximated by the analytical formalism in \citet{Inman2019PhRvD.100h3528I}:
\begin{equation}\label{eq:growth}
\begin{aligned}
   & D_{\rm PBH}(a)\approx \left(1+\frac{3\gamma a}{2a_{-} a_{\rm eq}}\right)^{a_{-}}\ ,
    \gamma=\frac{\Omega_{m}-\Omega_{b}}{\Omega_{m}}\ ,  \\ & a_{-}=\frac{1}{4}\left(\sqrt{1+24\gamma}-1\right)\ \ ,
\end{aligned}
\end{equation}
with $a_{\rm eq} \sim 1/3400$ representing the scale factor at matter-radiation equality. 

For the correlation between adiabatic and isocurvature modes, given by $P_{\rm corr}(k)$, we adopt the heuristic formula initially developed by \cite{Boyuan2022MNRAS.514.2376L} and updated in \citet{Boyuan2023arXiv231204085L}, applied to the range $ k_{\rm eq} < k < 3 k_{\mathrm{PBH}}$:
 \begin{equation}\label{eq:mixing}
\begin{aligned}
    P_{\rm corr}(k) = f_{\mathrm{PBH}} D_0\left(k/ k_{\mathrm{PBH}}\right)^3 P_{\rm ad}(k).
\end{aligned} 
\end{equation}
 Further work will be needed to fully explore this heuristic term\footnote{We have also compared the calculated power spectrum from Equ.~(\ref{eq:pk}) with that directly measured from the initial condition of our simulation box for both $\Lambda$CDM and PBH universes. The measurement for the $\Lambda$CDM simulation box is in good agreement with the \textit{Planck15} spectrum \citep{Plank2016A&A...594A..13P} at $k \sim 10-10^3 \,h\,\mathrm{Mpc}^{-1} $. For the PBH simulation boxes, a reasonably good convergence was found on small scales for all cases at around $k \gtrsim 100\,h\,\mathrm{Mpc}^{-1}$ , which corresponds to the isocurvature term $P_{\rm iso}(k)$ in Equ.~(\ref{eq:pk}). On larger scales ($k \lesssim 100\,h\,\mathrm{Mpc}^{-1}$), the measured power spectrum starts to deviate from the analytical form. Besides, limited by the box size, the contributions of large-scale modes $k \ll 10 \,h\,\mathrm{Mpc}^{-1} $ are missing. We conclude that due to the nonlinear effects around individual PBHs, the heuristic term $P_{\rm corr}(k)$ in Equ.~(\ref{eq:mixing}) might not accurately capture the perturbations by PBHs. To find a more accurate and general analytical form of $P_{\rm corr}(k)$, we need larger simulation boxes with higher resolution, which warrants future work.}. 

\begin{figure*}[!htb]
    \centering
\includegraphics[width= \linewidth]{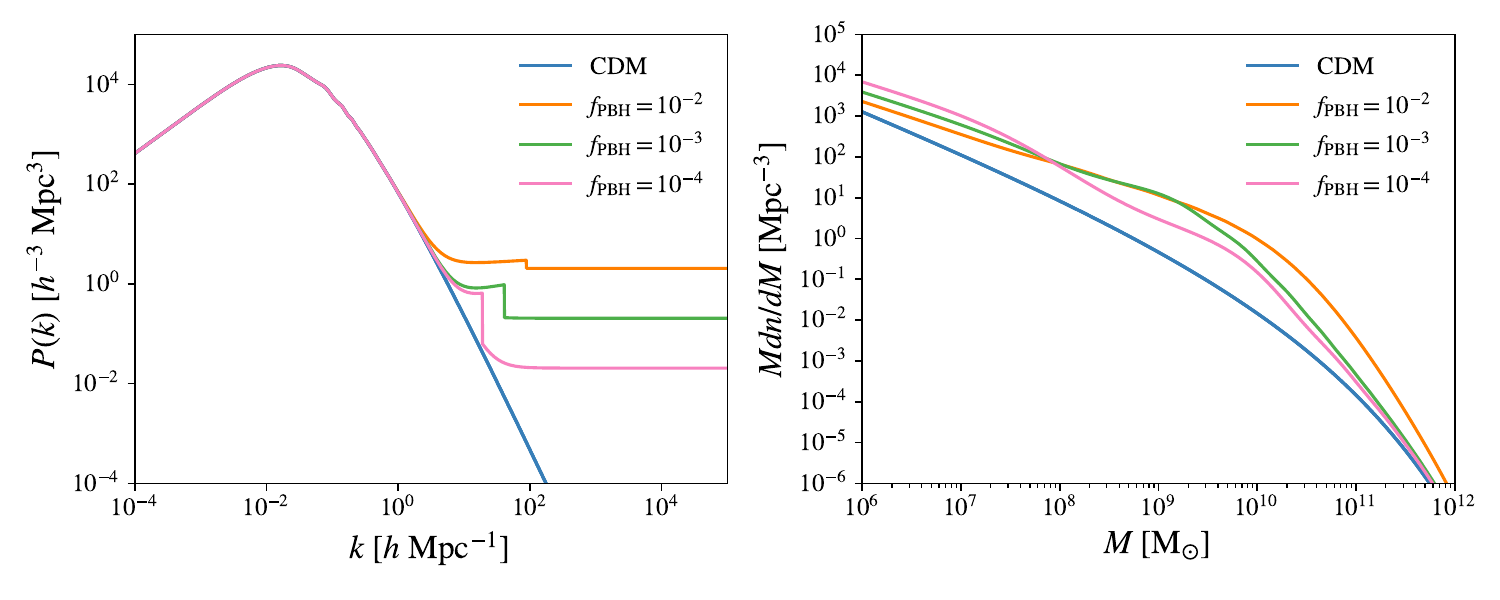}
\vspace{-25pt}
\caption{Structure formation within PBH cosmologies. {\it Left panel:} Power spectrum of the DM density field for different PBH mass fractions, $f_{\rm PBH}=10^{-4}$ (pink), $10^{-3}$ (green), 0.01 (orange), calculated following Equ.~(\ref{eq:pk})-(\ref{eq:mixing}), in comparison with the standard $\Lambda$CDM case (blue) measured by \citet{Plank2016A&A...594A..13P}. {\it Right panel:} The analytical HMF at $z=10$, employing the same conventions for the line color as before. It is evident that the excess power on small scales boosts the number density of low-mass DM halos, when PBHs are present. }
    \label{fig:pkHMF}
\end{figure*}

By convolving the total power spectrum with a top-hat filter, we apply the PS formalism \citep{Press1974ApJ} to compute the HMF, $dn/dM_{\rm h}$, incorporating corrections for ellipsoidal collapse \citep{Sheth1999MNRAS.308..119S}. In Figure~\ref{fig:pkHMF}, we show the total power spectrum including the PBH isocuvature and correlation terms, calculated with Equ.~(\ref{eq:pk})-(\ref{eq:mixing}), together with the resulting halo mass function as a function of $\fPBH$. At the lower mass end of the HMF, our analytical formalism exhibits a more complex pattern, while at the high mass end, it asymptotically approaches the $\Lambda$CDM limit with smaller $\fPBH$. 

\section{Numerical Methodology}\label{sec:sim}

\subsection{Simulation Box and Initial Parameters}

In this work, we focus on the formation of large-scale structure within PBH cosmologies, limiting our simulations to dark matter (DM) only\footnote{Accordingly, we set $\Omega_{\rm b} = 0$ in Equ.~(\ref{eq:growth}) to calculate the growth factor. This causes a small difference in the HMF derived from the PS formalism compared with the case using the \textit{Planck15} value $\Omega_{\rm b}=0.0486$ \citep{Plank2016A&A...594A..13P}.}.
Consequently, we defer the more complex simulations involving baryonic star formation and feedback physics to future studies. Our simulations are conducted using the state-of-the-art simulation code \textsc{gizmo} \citep{Hopkins2016MNRAS.455...51H}, incorporating the Tree+PM gravity solver from \textsc{gadget-3} \citep{springel2005cosmological}. In total, we have executed 8 simulations under various PBH initial configurations, as detailed in Table \ref{Table:SimParam}.

\begin{table*}[ht!]
    \centering
    \caption{Summary of key properties and parameters for the eight simulations. $m_{\rm PBH}$ represents the initial PBH mass (assuming all PBHs have the same initial mass), set at $10^6\,\Msun$ throughout. $f_{\rm PBH}$ denotes the mass fraction of PBHs relative to the total DM content. $N_{\rm gen}$ indicates the number of PBHs generated from the initial conditions. NONLIN is a flag indicating whether the initial non-linear perturbations of PDM particles by PBHs are included (\cmark) or not (\xmark). Similarly, CORR is a flag indicating whether the growth factor of PBH induced perturbations is corrected to reproduce the 
    growth of halos seeded by individual PBHs from the spherical collapse theory \citep{Mack2007ApJ}. 
    The final two parameters, $\epsilon_{\rm DM}$ and $\epsilon_{\rm PBH}$, represent the (co-moving) softening lengths for PDM and PBH particles, respectively.}
    \begin{tabular}{cccccccc}
    \hline
        Run  & $m_{\rm PBH}\ [\rm M_{\odot}]$  & $f_{\rm PBH}$  & $N_{\rm gen}$ & NONLIN & CORR  & $\epsilon_{\rm DM} [\rm kpc/h]$ & $\epsilon_{\rm PBH} [\rm kpc/h]$  \\
    \hline
        \texttt{CDM}  & - & - & - & - & - & 0.05 & -  \\
        \texttt{PBH2} & $10^{6}$ & $10^{-2}$ & 989 & \cmark & \cmark  & 0.05 & 0.002  \\
        \texttt{PBH3} & $10^{6}$ & $10^{-3}$  & 97 & \cmark & \cmark & 0.05 & 0.002  \\
        \texttt{PBH4} & $10^{6}$ & $10^{-4}$  & 10 & \cmark & \cmark & 0.05 & 0.002    \\
    \hline 
        \texttt{PBH3\_PS}  & $10^{6}$ & $10^{-3}$ & 97 & \xmark & \cmark & 0.05 & 0.002  \\
        \texttt{PBH3\_NS} & $10^{6}$ & $10^{-3}$  & 97 & \cmark & \xmark & 0.05 & 0.002  \\
        \texttt{PBH3\_PS\_NS} & $10^{6}$ & $10^{-3}$  & 97 & \xmark & \xmark & 0.05 & 0.002  \\
        \texttt{PBH3\_smooth} & $10^{6}$ & $10^{-3}$  & 97 & \cmark & \cmark & 0.05 & 0.5  \\
    \hline
    \end{tabular}
    \label{Table:SimParam}
\end{table*}

To select our simulation box, we consider a trade-off between resolution and statistical significance. We aim to generate a substantial number of PBHs within our box. Concurrently, we seek to resolve the forces between PDM and PBH particles, necessitating that the mass of PDM particles be significantly lower than that of PBH particles, thereby requiring high simulation resolution. Given the constraints of data storage and computational resources, we limit the total particle number to $N = 256^3$. For PBHs, we consider mass-dependent constraints on their abundance \citep[for a general review, see][]{Carr2021}. We adopt a fiducial value of $f_{\rm PBH} = 10^{-3}$ (\texttt{PBH3}), centered at $M_{\rm PBH} = 10^6\ \Msun$. To generate approximately $\mathcal{O}(10^2)$ PBHs within the box, we select a box size of $\sim 1\ \mathrm{(Mpc/h)}^3$ in comoving units. Consequently, the mass of PDM particles within this box is $\sim 5,100 \ \mathrm{M}_{\odot}/\mathrm{h}$, which also meets the force resolution requirement. We generate the initial conditions using the \textsc{MUSIC} code \citep{hahn2011multi} at an initial redshift of $z_{\rm ini} = 1/a_{\rm ini}-1 = 300$, where the growth of structure is still linear, and terminate our simulations at $z=10$, above which the box size is marginally representative. To facilitate a comprehensive comparison of the impact of PBHs on structure formation, we also consider the standard $\Lambda$CDM case (\texttt{CDM}) and other potential PBH fractions with the same PBH mass: $f_{\rm PBH} = 10^{-2}$ (\texttt{PBH2}), and $10^{-4}$ (\texttt{PBH4}).

In our simulations, we set the softening length for PDM particles to $\epsilon_{\rm DM} \sim 0.01 L/N^{1/3} \simeq 0.05\ \mathrm{kpc/h}$, emulating the collisionless nature of PDM. The softening length for PBHs is chosen to be $ \epsilon_{\rm PBH} \sim 0.002 \ \mathrm{kpc/h} $, simulating the dynamics of a point particle to better resolve the small-scale structure around PBHs. Additionally, we explore an extreme case for PBHs where their discrete nature is smoothed out (\texttt{PBH3\_smooth}), with a softening length for PBHs of $ \sim 0.5 \ \mathrm{kpc/h}$, to examine the influence of this parameter on structure formation.

\subsection{PBH initial conditions}
\label{sec:ic}
We establish the initial conditions for simulations including PBHs based on the analytical framework presented by \citet{Inman2019PhRvD.100h3528I}, and the numerical recipes developed in \citet{Boyuan2022MNRAS.514.2376L, Boyuan2023arXiv231204085L}, taking into account the density perturbations induced by PBHs across various scales. For the reader's convenience, we again briefly summarize the procedure \citep{Boyuan2022MNRAS.514.2376L, Boyuan2023arXiv231204085L}, reproducing key expressions from these studies.  

Initially, we add PBH particles into the $\Lambda$CDM cosmology initial conditions generated by \textsc{music} \citep{hahn2011multi}, according to the following algorithm:
\begin{itemize}
\item Divide the simulation box into $N_{\rm grid}$ grid cells, each containing approximately one PBH. 
\item Iterate over all cells. For the $i$-th cell, randomly draw a number $N_i$, following a Poisson distribution, as the total PBHs within the cell and randomly place them within the cell, noting their positions. Assign each PBH the velocity of its nearest neighbor PDM particle.
\item Reduce the mass of each PDM particle by a fraction $\simeq f_{\rm PBH}$ for mass conservation.
\end{itemize}

Next, we incorporate the isocurvature perturbation term induced by PBHs into the initial positions and velocities of PDM particles, denoted by $\left[T_{\rm iso}(a)-1\right]f_{\rm PBH} \delta_{\rm iso}^0$. Utilizing the Zel'dovich approximation \citep{Zeldovich1970A&A.....5...84Z}, we calculate the comoving displacement $\vec{\psi}$ and velocity $\Delta \mathbf{v}$ fields, resulting from the comoving acceleration field $\nabla \phi_{\rm iso}(\mathbf{x})$, as follows:
\begin{equation}
   \begin{aligned}
    \vec{\psi}(\vec{x})&=-\frac{2D_{\rm PBH}(a_{\rm ini})}{3\Omega_{m}H_{0}^{2}}\nabla \phi_{\rm iso}(\vec{x})\ , \quad
     \\ \Delta \vec{v}(\vec{x}) &=\frac{a_{\rm ini}\dot{D}_{\rm PBH}(a_{\rm ini})}{D_{\rm PBH}(a_{\rm ini})}\vec{\psi}(\vec{x})\,  \\
    \nabla \phi_{\rm iso}(\vec{x})&=4\uppi Gm_{\rm PBH}\sum_{i}\frac{\vec{x}-\vec{x}_{i}}{|\vec{x}-\vec{x}_{i}|^{3}}\,,
\end{aligned} \label{eq:field}
\end{equation}

where $\vec{x}_{i}$ represents the comoving coordinate of the $i$-th PBH at $a_{\rm ini}$, and the growth factor $D_{\rm PBH}(a_{\rm ini})$ is given by Equ.~(\ref{eq:growth}). The comoving acceleration field $\nabla \phi_{\rm iso}(\mathbf{x})$ accounts for contributions from all PBHs. Applying these equations, we adjust the positions and velocities of each PDM particle $j$ from its original coordinate $\Vec{x}_j$ and velocity $\Vec{v}_j$ in the $\Lambda$CDM initial conditions to $\hat{\Vec{x}}_j = \Vec{x}_j + \vec{\psi}(\vec{x}_j)$ and $\hat{\Vec{v}}_j = \Vec{v}_j + \Delta \Vec{v}(\vec{x}_j)$, respectively.

Given the `seed' effect \citep[as proposed by][]{Mack2007ApJ}{}, perturbations near PBHs are nonlinear from the outset. For simplicity, we define a nonlinear scale for a single PBH via $d_{\rm nl} = [D_{\rm PBH}(a_{\rm ini})/\bar{n}_{\rm PBH}]^{1/3}$, where $\bar{n}_{\rm PBH}$ is the average PBH number density. To consistently model both `seed' and `Poisson' effects, we introduce two numerical treatments. On scales significantly larger than $d_{\rm nl}$, we can effectively use linear approximation to the initial conditions, by applying a truncated displacement field as 
\begin{equation}
    \Vec{\psi}_{\rm Poisson} = \min (1, d_{\rm PDM}/|\Vec{\psi}|)\Vec{\psi},
\end{equation}
where $d_{\rm PDM}$ is the average distance between PDM particles as given by the initial resolution of the simulation.

In regions very close to PBHs ($r \lesssim d_{\rm nl}$), we account for nonlinear structures with stronger perturbations than allowed by truncation, as indicated by NONLIN in Table \ref{Table:SimParam}. Considering a virialized halo as the outcome of the `seed' effect with an overdensity $\Delta_{\rm vir} = 18\pi^2$, we apply a specific truncation to capture the nonlinear structure:
\begin{equation}
    \begin{aligned}
    \Vec{\psi}_{\rm seed} & = \min (1, f_{\rm shrink} r /|\Vec{\psi}|)\Vec{\psi}, \quad  \\ f_{\rm shrink} & = 1 - \Delta_{\rm vir}^{-1/3} \approx 0.822.
\end{aligned}
\end{equation}
Here, $r$ is the distance of the PDM particle from its nearest PBH. With this prescription, we create a uniform sphere of PDM surrounding the PBH with an overdensity of $\Delta_{\rm vir}$ and a radius $d_{\rm nl}$, to approximate the nonlinear features. By running our \textsc{gizmo} simulations (see the next section), we find that the mass of the virialized halo around an isolated PBH grows linearly  with $a$, as $M_{\rm h}\sim (a/a_{\rm ini})(4\pi/3)D_{\rm PBH}(a_{\rm ini})m_{\rm PBH}$, 
which is generally consistent with the spherical collapse theory~\citep[see][]{mo2010galaxy}{}{}, where $M_{\rm h} \sim m_{\rm PBH} a/a_{\rm eq}$ \citep{Mack2007ApJ}, but the normalization is off by a factor of a few. To mitigate the difference in halo mass between the analytical predictions by \citet{Mack2007ApJ} and our simulations using the growth factor from Equ.~(\ref{eq:growth}), we correct the growth factor to $D_{\rm PBH}(a_{\rm ini}) \sim 3 a_{\rm ini} / (4\pi a_{\rm eq})$, as indicated by CORR in Table \ref{Table:SimParam}. Finally, we combine the two perturbation regimes to calculate the displacement field induced by PBHs as $\Vec{\psi} = \Vec{\psi}_{\rm seed} +\Vec{\psi}_{\rm Poisson} $, capturing both `seed' and `Poisson' effects.

For completeness, we conduct simulations with the same PBH parameters but without correcting the growth factor (\texttt{PBH3\_NS}), without the seed effect (\texttt{PBH3\_PS}), and without both (\texttt{PBH3\_PS\_NS})\footnote{Specifically, when the NONLIN flag is turned on, the Poisson term $\Vec{\psi}_{\rm Poisson}$ include the contributions from all PBHs but the nearest one, while the seed term $\Vec{\psi}_{\rm seed}$ comes from the nearest PBH. If NONLIN is turned off, we only consider the Poisson term contributed by all PBHs.}. The latter case replicates our prescription from previous work \citep{Boyuan2022MNRAS.514.2376L}.

\section{Results and Discussion}\label{sec:result}
\subsection{Halo Statistics in PBH cosmologies}
To derive the HMF from simulation snapshots, we rely on the \href{https://bitbucket.org/gfcstanford/rockstar/src}{\textsc{rockstar}} halo finder \citep{behroozi2012rockstar} to assign particles to halos. However, one caveat is that when reading the snapshot files, \textsc{rockstar} only considers PDM particles in calculating the halo mass. Therefore, we need to manually assign the mass of PBHs to their respective host halos in our calculations. To exclude less massive halos that contain too few PDM particles and are thus poorly resolved dynamically, we only consider halos above the molecular cooling threshold, set by $M_{\rm h} \gtrsim 1.5 \times 10^5\,\Msun \left(\frac{1+z}{31}\right)^{-2.1}$ \citep{Trenti2009ApJ...694..879T}. Given that this mass is typically much smaller than the halos seeded by PBHs in the redshift range of $z\simeq 10-30$, we employ it as an approximate lower mass limit for PDM-only halos\footnote{Previously, we derived the star-forming minihalo mass threshold within the stellar-mass ($10-100\ \rm M_\odot$) PBH regime by comparing the molecular/radiative cooling vs. the heating rates by PBH accretion \citep{Boyuan2022MNRAS.514.2376L}. Calculating the precise lower mass limit for the host halos where first star formation is triggered under the influence of PBH accretion feedback within other PBH mass regimes will require hydrodynamical simulations to assess cooling and heating, which is beyond the scope of this study.}, defined as halos containing no PBHs. 

Initially, when analyzing the PBH simulations, we observed that halos are positioned closer together, with some overlapping at small scales, compared to the morphology in the \texttt{CDM} (pure PDM) simulation. At the small scales involved, baryonic effects will be important, in particular the feedback from star formation \citep[e.g.,][]{Brooks2013ApJ...765...22B, Jaacks2019}. As simulations including baryonic feedback are beyond the scope of our current study, and the numerical prescriptions used to produce cosmological initial conditions including PBHs are imperfect \citep[][see the discussion in their Sec.~2.2]{Boyuan2023arXiv231204085L}, we cannot accurately account for the statistics of subhalos, and we defer a more detailed analysis of subhalo structure within PBH cosmologies to future work. For simplicity, we therefore remove all subhalos from our halo catalog, retaining only their most massive hosts. Next, we summarize our algorithm to address this cloud-in-cloud problem and to remove subhalos:

\begin{itemize}
    \item[1.] Iterate over each PBH in the PBH catalogue, indexed by $i$. For the $i$-th PBH, iterate over the halo catalog indexed by $j$. Identify and record the index of the most massive host halo by calculating the Euclidean distance from this PBH to the center of the $j$-th halo and comparing it with the virial radius of the $j$-th halo.
    \item[2.] Iterate over the halo catalog indexed by $j$. For each halo that is not in the local maximum-mass host list, calculate the Euclidean distance between the center of $j$-th halo and the center of the local maximum host. Add this halo to the removal list, if this separation distance is smaller than the virial radius of the local maximum host.
    \item[3.] Iterate over PBHs, indexed by $i$, to find the index of their most massive hosts. Add the PBH mass to these most massive hosts. 
    \item[4.] Calculate the pairwise separations among the remaining halos that are not local maximum hosts. For each pair, the less massive halo is removed if the separation is smaller than the virial radius of the more massive halo.
\end{itemize}

On the other hand, for the \texttt{CDM} run, we only apply step 4 to remove all subhalos from the halo catalogue, to facilitate a meaningful comparison with the PBH cases. This method allows us to systematically exclude all subhalos within each host halo. Clearly, developing a halo finder capable of properly capturing the subtleties introduced by PBHs in cosmic structure formation is an important goal for future work.


\subsection{Large-Scale Structure}

\begin{figure*}[ht!]
    \centering
    \includegraphics[width=  \linewidth]{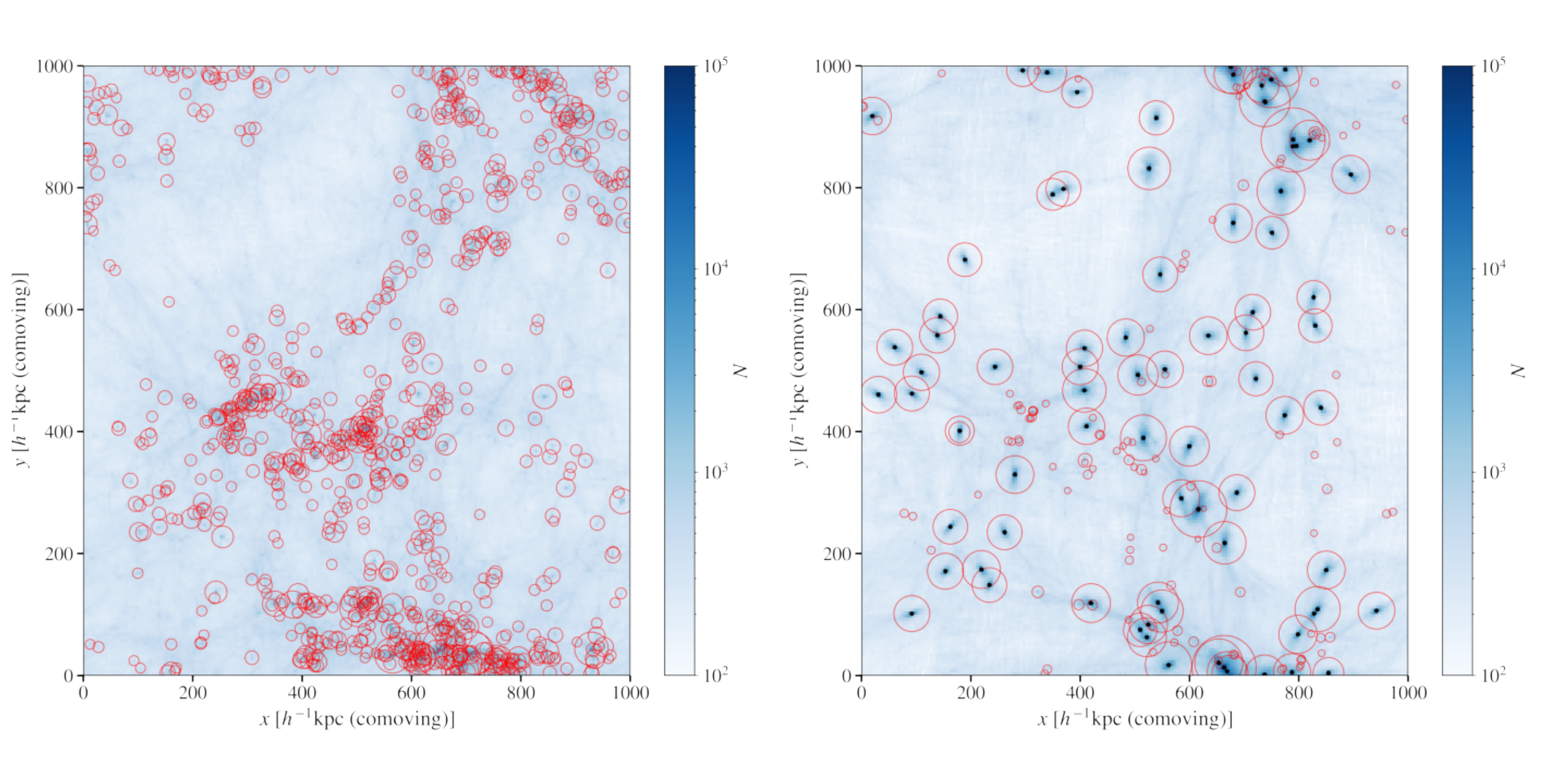}
    \vspace{-25pt}
    \caption{Distribution of matter and PBHs. We project the simulation box onto a 2D plane to depict the cosmic web at $z=10$ (in comoving units). {\it Left panel:} Cosmic web in the baseline \texttt{CDM} simulation. {\it Right panel:} Distribution for the \texttt{PBH3} run, with $f_{\rm PBH} = 10^{-3}$. In this visualization, black dots represent the positions of PBHs, while red circles indicate the positions of halos, with their radii corresponding to the comoving virial radii of the halos. The color bar denotes the number of PDM particles stacked within each pixel. In the right panel, all subhalos within their more massive hosts have been removed to better showcase the spatial distribution of halos and PBHs.}
    \label{fig:cswb}
\end{figure*}

\begin{figure*}[!htb]
    \centering
\includegraphics[width=0.9 \linewidth]{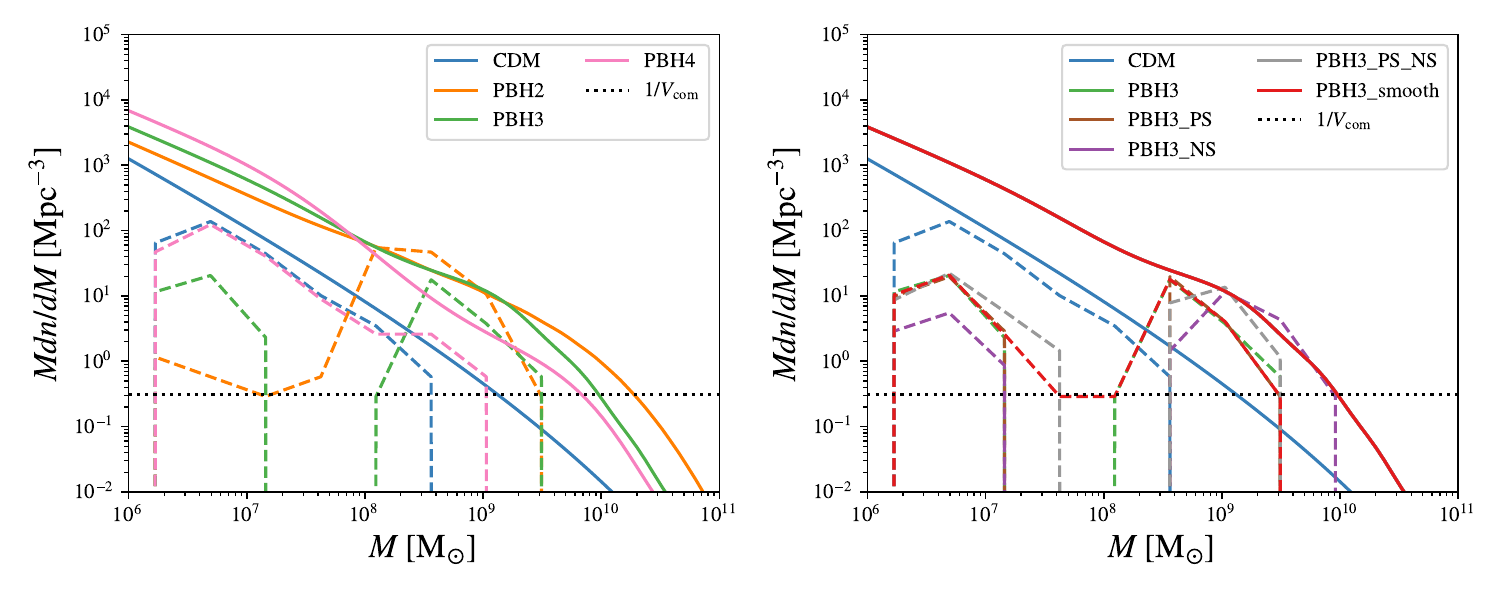}
\vspace{-15pt}
\caption{Comparing the simulated halo mass function (dashed lines) from different initial conditions after removing all subhalos at $z=10$. To provide a reference point, we also reproduce the HMF derived from the analytical formalism in Section~2 (solid lines) at the same redshift, noting that the same line color indicates identical initial conditions and PBH parameter. {\it Left panel:} Effect of varying the fraction of PBHs in the DM, specifically considering $f_{\rm PBH} = 10^{-2}$ (\texttt{PBH2}), $10^{-3}$ (\texttt{PBH3}), and $10^{-4}$ (\texttt{PBH4}). {\it Right panel:} Effect of varying numerical implementation and parameters, for a fixed $f_{\rm PBH} = 10^{-3}$. Specifically, we consider the absence of the correction of the growth factor (\texttt{PBH\_NS}), non-linear perturbations from the nearest PBH (\texttt{PBH\_PS}), and of both (\texttt{PBH\_PS\_NS}). Additionally, the softening parameter for PBH particles is increased in the \texttt{PBH3\_smooth} case. In both panels, the horizontal dashed line represents the number density resolution limit set by the comoving volume of our box as $1/V_{\rm com}$. }
    \label{fig:HMF}
\end{figure*}


After removing all subhalos, we generate projection plots at the last snapshot of the \texttt{CDM} and \texttt{PBH3} simulations at $z=10$, depicted in Figure~\ref{fig:cswb}. These plots illustrate the structural differences when PBHs constitute $\fPBH = 10^{-3}$ of the total DM, compared to a standard $\Lambda$CDM simulation. On a larger scale, more massive halos have formed by this redshift in the \texttt{PBH3} simulation, while significantly fewer have formed in the \texttt{CDM} case. Additionally, the \texttt{PBH3} run demonstrates clear segregation in halo mass, with almost all massive halos containing PBHs and being substantially larger than their pure PDM halo counterparts. Furthermore, the abundance of less massive halos is suppressed compared to the \texttt{CDM} run. Structurally, the average spacing between halos is larger in the \texttt{PBH3} case. We also examined these effects in the \texttt{PBH2} and \texttt{PBH4} runs to understand how PBH abundance impacts structure formation. Surprisingly, almost all halos in the \texttt{PBH2} run contain at least one PBH, and pure PDM halos have almost vanished. The \texttt{PBH4} simulation, however, more closely resembles the \texttt{CDM} one, displaying a mix of clumpy pure PDM ($\sim$$10^6\,\Msun$) minihalos and a few PBH-hosting halos. For further details on how different PBH mass fractions $\fPBH$ influence structure formation, we refer the reader to Figure~\ref{fig:cswb_2} in Appendix~\ref{AppendA}.

Subsequently, we count the halos within each mass bin and plot the HMF after subhalo removal at $z=10$, as shown in Figure~\ref{fig:HMF}\footnote{We have also checked the convergence of the HMF from simulations with the analytical form. General convergence is observed at around $\mathcal{O}(10^8) \Msun$, and the simulated HMF shows some deviation at low masses. This difference can be explained by the non-linear effects induced by PBHs at small scales, which cannot be fully modelled by the PS formalism. Here, to show the effect of PBH seeding and bimodal features in halo mass distributions, we only plot the mass distribution after subhalo removal. }. For ease of comparison, we also reproduce the analytical HMFs, as derived in Section~\ref{sec:hmf}\footnote{The analytical HMFs displayed in Figure~\ref{fig:HMF} are zoomed-in versions of those shown in the right panel in Figure~\ref{fig:pkHMF}, selecting the most relevant mass scale from our simulations for better visualization purposes.}. The left panel of Figure \ref{fig:HMF} illustrates the effects of varying the PBH mass abundance parameter $\fPBH$, and the right panel displays the effects of different initial conditions or numerical implementations for the same $\fPBH$. Here, all simulations are represented with dashed lines. The sharp cutoff in our simulated results can be attributed to the halo selection mass limit at the lower end and the box size limit at the higher end. From the left panel, we find that with higher $\fPBH$, the abundance of halos with mass $\ll 10^8 \Msun$ is greatly suppressed, in contrast to the analytical results. Comparing the simulation results in the right-hand panel, the differences are minimal, indicating that variations in the choice of numerical parameters do not greatly change the HMF. However, in the \texttt{PBH3\_NS} implementation, we observe that the mass of the most massive halos is approximately twice as large as in the \texttt{PBH3} run. This effect is likely due to the larger structure growth factor $D_{\rm PBH}$ without the correction motivated by the results in \citet[][see Sec.~\ref{sec:ic}]{Mack2007ApJ}, resulting in heavier PBH-seeded halos. At the lower mass end ($\ll 10^8\Msun$), the halo abundance in the \texttt{PBH3\_NS} run is suppressed by a factor of 2 as compared to the \texttt{PBH3} case. Both panels suggest that more massive halos suppress the abundance of less massive halos, a topic we will elaborate on in more detail in Section \ref{subsec:NN}.

\subsection{Small-Scale Structure}

\begin{figure*}[ht!]
    \centering
    \includegraphics[width=  \linewidth]{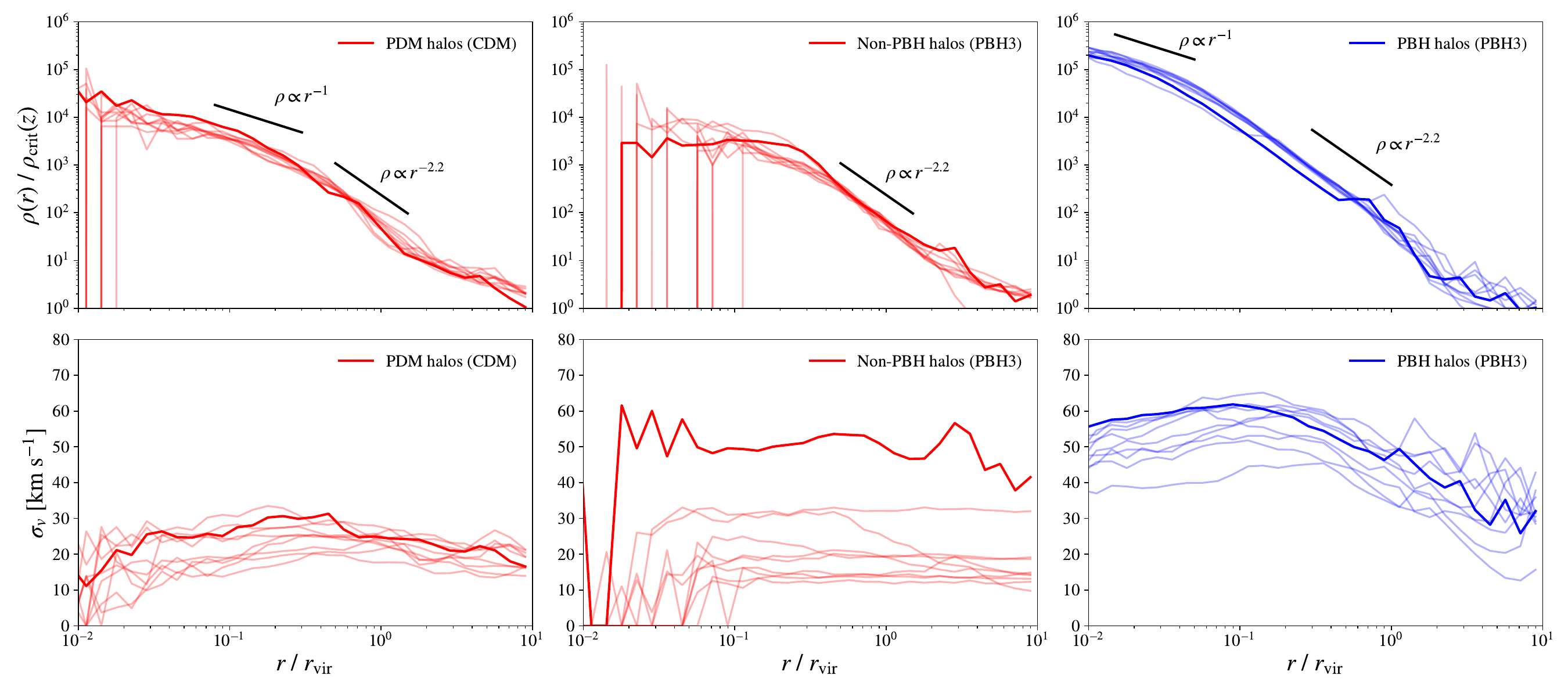}
    \vspace{-25pt}
    \caption{Structure of massive halos in the \texttt{CDM} and \texttt{PBH3} runs at $z=10$. From the \texttt{PBH3} run, we specifically select the ten most massive halos that do not contain PBHs (red, {\it mid panels}), alongside with the ten most massive PBH hosting halos (blue, {\it right panels}). For comparison, we also select 10 most massive halos from the \texttt{CDM} run without any influences from PBHs (red, {\it left panels}). The opaque curves represent the most massive halos among the selected ones, whereas the others are represented by quasi-transparent curves. {\it Top panels:} Halo density profile. Here, we normalize $\rho(r)$ to the average cosmic density $\rho_{\rm crit}(z)$ at $z=10$, and represent the physical distance to a halo's center, $r$, normalized by the halo's virial radius ($r_{\rm vir}$). {\it Bottom panels:} Velocity dispersion profile, given by $\sigma_v$ vs. $r/r_{\rm vir}$.}
    \label{fig:rhohalo}
\end{figure*}

Among the halo catalogue for the \texttt{PBH3} run, we find that almost all of the more massive halos ($\gtrsim 10^8\,\Msun
$) host PBHs, with only one being a pure PDM halo. To fairly compare PBH hosting halos and pure PDM ones, we select the ten most massive pure PDM halos with mass range of $\sim 3\times 10^6 - 2\times 10^8 ~\Msun $ and compare them with the ten most massive PBH host halos ranging from $\sim 3-9\times 10^8 ~\Msun$. In this way, the best resolved PDM halos are selected for comparison, as seen from Figure\, \ref{fig:rhohalo}. For comparison, we also select the ten most massive PDM halos with mass range of $\sim 4 \times 10^7 - 1\times 10^8 ~\Msun$ from the \texttt{CDM} run as the simulation without PBH. Our results are shown in Figure \ref{fig:rhohalo}. We observe that a few selected PDM halos exhibit a constant density near the center, forming a core-like profile. These core-flattened halos are interacting with the nearest more massive PBH host halos, resulting in perturbed density profiles. 


The remaining PDM halos in the \texttt{PBH3} run are less resolved due to their low masses close to the resolution limit, whereas PBH host halos typically exhibit $\rho \propto r^{-1}$ scaling for the central profile. At larger distances from the halo center, both PBH host and PDM halos exhibit a steeper profile, with $\rho \propto r^{-2.2}$, largely consistent with the Navarro-Frenk-White (NFW) profile. A similar scaling is also observed when sampling PDM halos in the \texttt{CDM} run for comparison. We also examine the velocity dispersion versus normalized radius in the right panel of Figure~\ref{fig:rhohalo}. We find that the overall velocity dispersion is higher for PBH hosts, as the selected cases are more massive, and we note that PDM halos show a more flattened radial profile. 

Given that this work involves a DM-only simulation,  whereas baryonic feedback is known to dominate on small scales \citep[e.g.,][]{Brooks2013ApJ...765...22B, Wetzel2016}, we defer further studies involving baryonic physics to future work in order to better simulate smaller-scale structure and stellar feedback processes within PBH cosmologies.

\subsection{Nearest-Neighbor Statistics}\label{subsec:NN}

\begin{figure*}[ht!]
    \hspace{-2cm}
    \centering
    \includegraphics[width= \linewidth]{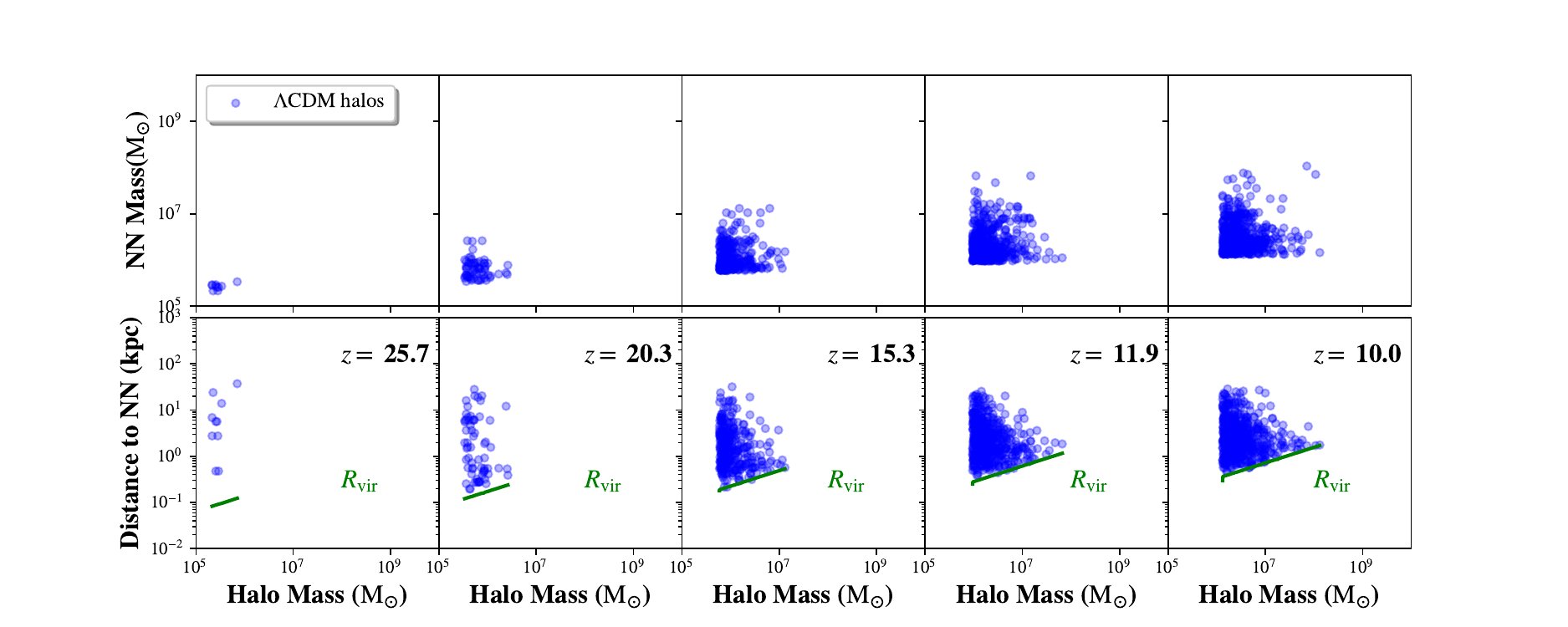}
    \vspace{-10pt}
    \caption{Nearest-neighbor (NN) statistics for standard $\Lambda$CDM case, represented by our \texttt{CDM} simulation. We show the relationship between halo mass and the mass of its NN halo ({\it top panels}), as well as the physical distance to the NN ({\it bottom panels}), for all halos in the simulation box (represented by blue circles), over $z \simeq 25 - 10$. Additionally, we indicate the virial radius for different halo masses ({\it green lines}). For snapshots larger than a redshift of $\sim 25$, the scarcity of halos precludes the generation of such plots.}
    \label{fig:NNCDM}
\end{figure*}

\begin{figure*}[ht!]
    \hspace{-2cm}
    \includegraphics[width=1.2\textwidth]{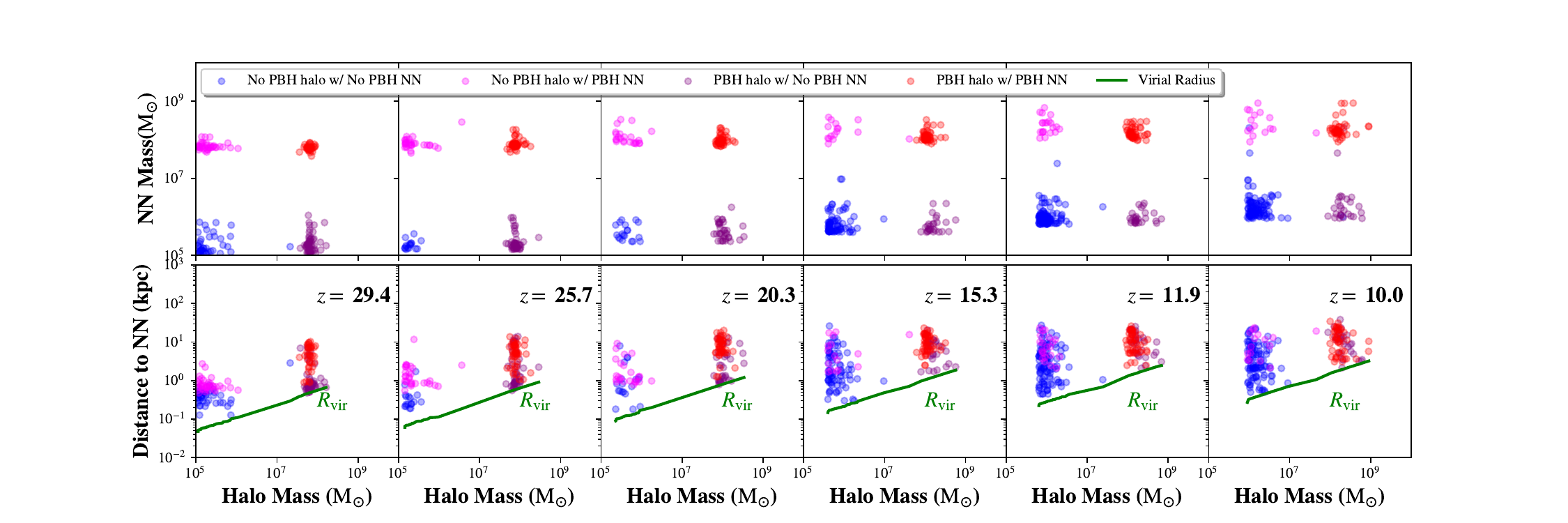}
    \vspace{-20pt}
    \caption{NN analysis for the \texttt{PBH3} simulation, where the PBH density fraction is $\fPBH = 10^{-3}$. The same conventions are employed as in Fig.~\ref{fig:NNCDM}. Here, four distinct marker colors are used to differentiate between halos and their NNs based on their association with PBHs: No-PBH halo with No-PBH halo NN (blue), No-PBH halo with a PBH halo NN (pink), PBH halo with No-PBH halo NN (purple), and PBH halo with a PBH NN halo (red). The distinction between the halos that contain PBHs and the pure PDM ones is clearly visible.}
    \label{fig:NNPBH3}
\end{figure*}

Next, we explore a possible way to explain key features in the HMF using nearest-neighbor (NN) statistics, an attractive choice due to its minimal parameter dependence, as previously explored by~\cite{Banerjee2021MNRAS.500.5479B} in the context of large-scale sky surveys. We plot the halo mass against the mass of, and distance to, its nearest neighbor halo in Figures~\ref{fig:NNCDM} and \ref{fig:NNPBH3}, representing
the \texttt{CDM} and \texttt{PBH3} cases, respectively, to examine the clustering patterns and their evolution through cosmic time. 

In the \texttt{CDM} run (Figure~\ref{fig:NNCDM}), we observe a hierarchical merger pattern, with most of the minihalos with $\sim 10^6\,\Msun$ forming near each other and merging into larger halos with $\gtrsim 10^8\,\Msun$ as they evolve with redshift. In contrast, in the \texttt{PBH3} run, we categorize halos into four distinct NN pairings: PDM halo with PDM halo, PDM halo with PBH host halo, PBH host halo with PDM halo and PBH host halo with PBH host halo. The top panels of Figure \ref{fig:NNPBH3} show a clustering pattern that indicates the distinction between PDM halos and PBH host halos. The time evolution of the NN distances show a more complex pattern, as the number of PDM halos paired with PBH host halos and vice versa decreases with decreasing redshift.

Further details on NN statistics for additional simulations, exploring our suite of parameters and initial setups, can be found in Appendix~\ref{AppendA}. NN plots for the \texttt{PBH\_PS} and \texttt{PBH2} runs (Figures \ref{fig:NNPBH_PS} and \ref{fig:NNPBH2} in Appendix~\ref{AppendA}) exhibit a consistent evolutionary pattern. However, in the \texttt{PBH4} run, PBH host halo pairings with other PBH host halos vanish as redshift decreases, indicating that all PBH halos are surrounded by PDM halos and disruption of PDM halos by PBH host halos is less pronounced. For PBH abundances of $\fPBH \gg 10^{-3}$ almost all halos host PBHs, and PDM halos formed at earlier redshifts nearly vanish. Conversely, for $\fPBH \ll 10^{-3}$, the clustering pattern (Figure~\ref{fig:NNPBH_4}) is similar to that of the \texttt{CDM} run (Figure~\ref{fig:NNCDM}). Moreover, removing the `seed' effect in the initial condition does not significantly affect the clustering pattern, as shown in Figure~\ref{fig:NNPBH_PS}. 

\begin{figure*}[ht!]
    \centering
    \includegraphics[width=0.7\textwidth]{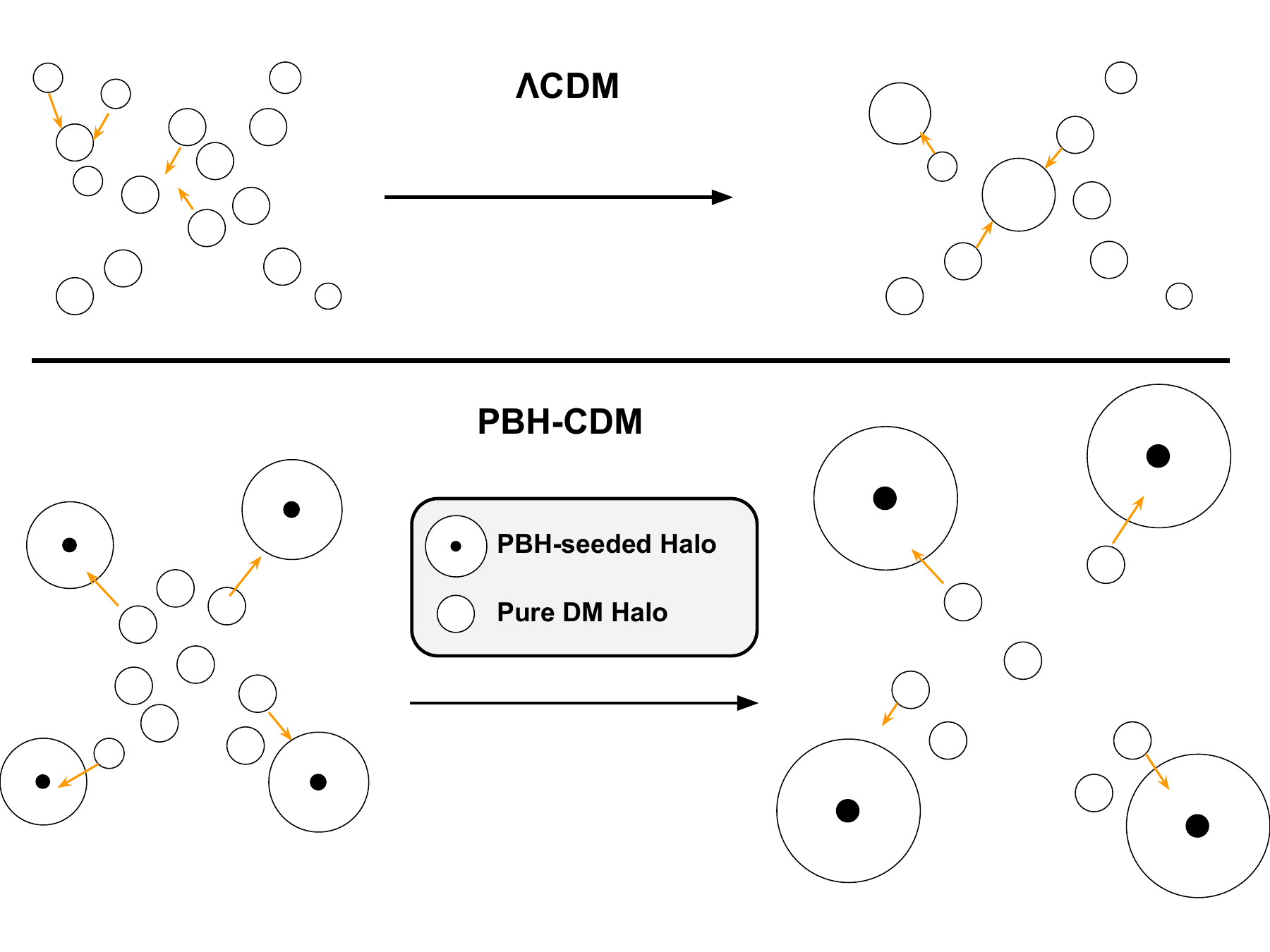}
    \vspace{-25pt}
    \caption{Halo assembly in the early universe. In the $\Lambda$CDM case, minihalos are expected to cluster and undergo hierarchical merging to form higher-mass halos ({\it top panel}). However, PBHs significantly disrupt this hierarchical merger process for minihalos formed close to PBH-seeded halos ({\it bottom panel}). With increasing PBH abundance, almost no minihalo can withstand this ingestion, and nearly all are likely to be absorbed by the nearest PBH-seeded halo. This dynamic interaction underscores the pivotal role of PBHs in altering the standard view of cosmic structure formation through their impact on hierarchical merging within the PBH-CDM setting.}
    \label{fig:halodyn_scheme}
\end{figure*}

\begin{figure*}[ht!]
    \centering
    \includegraphics[width=0.8\textwidth]{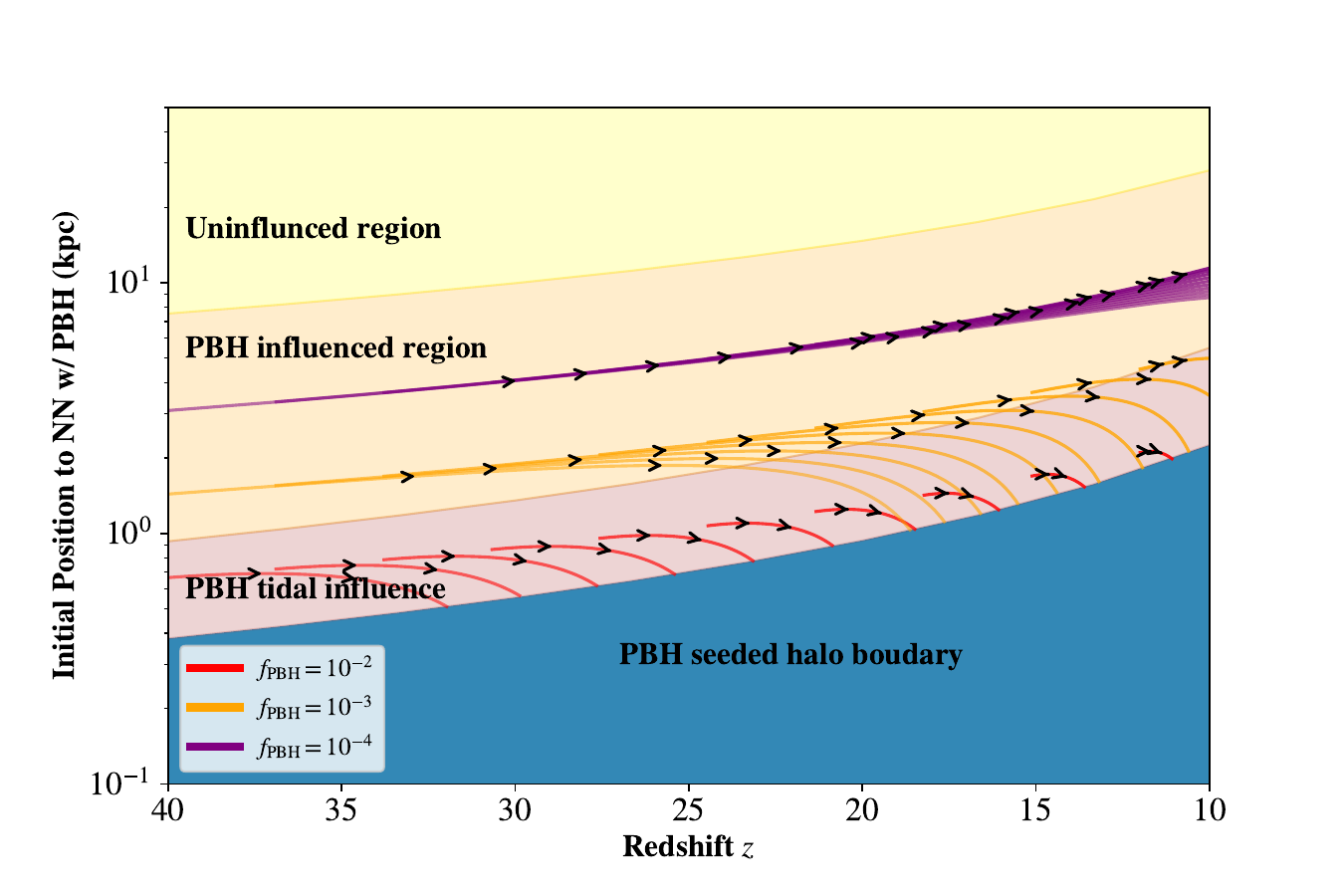}
    \vspace{-25pt}
    \caption{Schematic representation of halo dynamics within a PBH-CDM universe, modeling their trajectories relative to their nearest PBH-seeded halo. The vertical axis of `Initial Position' refers to the starting physical distance $r_{0}$ of a PDM halo in proximity to its closest PBH-seeded halo. 
  We show the regime where the trajectories of PDM halos are (not) affected by PBHs in yellow (light orange) according to Equ.~(\ref{eq:fate}).
    We further indicate the zones under PBH tidal influence (pink), and boundaries of isolated PBH-seeded halos (blue). 
    To illustrate the dependence of PDM halo dynamics on the PBH mass fraction $f_{\rm PBH}$, we plot exemplar trajectories of PDM halos starting from $r_{0}=0.2d_{\rm PBH}$
    for $f_{\rm PBH} = 10^{-2}$ (red), $f_{\rm PBH} = 10^{-3}$ (orange), and $f_{\rm PBH} = 10^{-4}$ (purple), indicating that more PDM halos will be disrupted/engulfed by PBH-seeded halos when $f_{\rm PBH}$ is higher.} 
    \label{fig:halodyn}
\end{figure*}

To further explore the phenomenology in our NN statistics, we consider an idealized dynamical model for the motion of PDM halos around nearby PBH-seeded halos
, to explain the bimodal feature in the halo mass function. We here specifically consider the competition between the self-gravity of a given region and the Hubble expansion in the presence of PBHs (schematically illustrated in Fig.~\ref{fig:halodyn_scheme}). For a PDM halo that forms at redshift $z_{0}$ 
at a physical distance $r_{0}$ from its NN PBH host halo, we assume that $r_{0}$ is sufficiently large for the two halos to be effectively treated as point masses. 

Further assuming a vanishing velocity at formation, the acceleration of a PDM halo at a proper distance $r$ from its NN PBH host halo, and at cosmic time $t$, can be effectively expressed as\footnote{For a similar argument, see Equ.~(5) in \cite{Ali-Haimoud2017PhysRevDMerger}, in the context of binary mergers.} 
\begin{equation}
    \frac{d^2 r}{dt^2} = r(H^2(t) + \dot{H}(t))- \frac{G\bar{M}_{\rm h, PBH}(t)}{r^2} \mbox{\ .}\label{eq:dyn}
\end{equation}
For simplicity, $\bar{M}_{\rm h, PBH}(t)$, the average mass of the PBH-seeded NN halo is approximated by $\bar{M}_{\rm h, PBH} \sim m_{\rm PBH} a/a_{\rm eq}$, based on the spherical collapse theory involving an isolated PBH \citep{Mack2007ApJ}. 
Numerically solving Equ.~(\ref{eq:dyn}), we identify a critical initial physical distance, beyond which the trajectory of a newly-formed halo is hardly affected by the presence of PBHs, implying that its merger and accretion history is not impacted as well, approximately given by  $r_{\rm crit} \simeq 30\, {\rm kpc}\,[(1+z_{0})/11]^{-1} $.

 We first compare this distance to the Hill radius \citep{Shu1982phyn.book.....S}, approximated as $R_{\rm Hill} = 2.4 R_{\rm vir}(\bar{M}_{\rm h, PBH}) \simeq 5.5\,{\rm kpc}\,[(1+z_0)/11)]^{-4/3} $, which represents the maximum radius for tidal influence from the PBH host halos. Since the tidal influence sphere is much smaller than the region (defined by $r_{\rm crit}$) where a PBH-seeded halo can disrupt the merger and accretion histories of its neighbors, the latter effect dominates at larger scales in the formation and growth of halos leading to a bimodal feature in the HMF. 
Therefore, the fate of a PDM halo can be told by comparing $r_{0}$ with $r_{\rm crit}$: 
\begin{equation}
r_{0} \begin{cases} \gg r_{\rm crit}: &\text{PBH non-influenced region} \\  \ll r_{\rm crit}: &    \text{PBH influenced region,}\end{cases}\label{eq:fate}
\end{equation}

i.e., $r_{\rm crit}$ divides the merger histories of PDM halos into those that will be significantly altered (influenced) by the presence of a nearby PBH, and those that will not.

 Therefore, this critical radius for PBH influenced regions sets a bifurcation limit for the inter-PBH separation and thus the mass fraction of PBHs within DM. If, on average, PBHs are too closely spaced, this will result in overlaps of the perturbation regions. In this case, trajectories of almost all PDM halos will be impacted. 
 Equating the critical radius $r_{\rm crit}$ to half the inter-PBH separation (proper) distance $d_{\rm PBH}\simeq (1/\bar{n}_{\mathrm{PBH}})^{1/3}(1+z)^{-1} \simeq  27 \,{\rm kpc}\, [(1+z_0)/11]^{-1} [f_{\rm PBH}/10^{-3}]^{-1/3}$, we calculate a rough bifurcation threshold for the PBH abundance (independent of $z_0$) as $\fPBH \simeq  10^{-4}$. If the PBH abundance is much larger than this limit, almost no PDM halos will survive if formed close to a PBH, and halos that have already formed will be engulfed by the NN PBH host halo. Conversely, if $\fPBH \ll 10^{-4}$, the existing halos will be much less disrupted, which facilitates the formation of more massive PDM halos through mergers with other nearby PDM halos. We illustrate our model in Figure~\ref{fig:halodyn}, showing different evolutionary possibilities. 

We again emphasize that this is an extremely simplified dynamical model, serving to illustrate the key governing physics in the complexity of our simulations, where PBHs interact with standard PDM.

\section{Summary and Conclusions}\label{sec:conclusion}

In this paper, we investigate how the presence of million-solar-mass PBHs impact the assembly history of DM halos in the early universe. Building on our earlier work to implement PBH initial conditions  \citep{Boyuan2022MNRAS.514.2376L,Boyuan2023arXiv231204085L}, we carry out DM-only cosmological simulations to study the impact of such $10^6\,\Msun$ PBHs on the HMF and early structure formation. Our primary findings indicate that PBHs can seed the formation of massive halos. When PBHs constitute a fraction $\fPBH \gg 10^{-4}$ of DM, the PBH-seeded halos efficiently disrupt and swallow their pure PDM-halo counterparts, leading to a bimodal feature in halo mass distribution. Additionally, if $\fPBH \gg 10^{-3}$, nearly all halos will contain at least one PBH, with PDM halos being engulfed by PBH-seeded halos shortly after their formation. When varying the initial conditions in terms of perturbation strength while maintaining the same PBH mass abundance, we find that these changes have non-trivial effects, but do not affect the bimodal feature in the HMF, compared to simulations that vary $\fPBH$. 

In modeling this halo formation scenario, we employ NN statistics to observe the clustering patterns of neighboring halo pairs and their evolution across redshift. To facilitate the physical interpretation, we develop a halo free-fall model to elucidate the evolution of structure and clustering patterns observed in the NN analysis. We identify a critical distance below which PDM halo formation is impacted, and any already formed PDM halo located much closer to its NN PBH-seeded halo than this distance will be engulfed. Furthermore, the critical distance can be compared with the inter-PBH separation distance $d_{\rm PBH}$, which gives a critical value for the PBH DM-fraction of $\fPBH \simeq 10^{-4}$, to ascertain whether we are entering a universe dominated by PBH-seeded halos.

Regarding structure formation in this PBH-CDM setting, we use the Press-Schechter (PS) formalism \citep{Press1974ApJ, Sheth1999MNRAS.308..119S} as an approximate avenue towards deriving the HMF analytically, assuming a Gaussian distribution of overdensity peaks. However, the overdensity from PBHs, due to their discreteness, takes the form of a delta function in real space, and the PBHs are Poisson distributed at small scales. These features evade the Gaussian assumption, so PBHs and their perturbations contribute non-trivially to the mass variance when taking the volume average over the window function \citep{Kashlinsky2016ApJ...823L..25K,Desjacques2018PhRvD..98l3533D,Deluca2020JCAP...11..028D}. The standard PS formalism may thus not adequately describe structure formation or the HMF, due to the deviation from Gaussian statistics in PBH cosmologies. One consequence is that the PS formalism does not sufficiently explain the predicted bimodal feature in the HMF. Future work, by carefully revisiting the evaluation of the filtered-mass variance $\sigma(M)$ and the nonlinear growth of structure around PBHs, needs to address these shortcomings. 

We recognize several limitations in our study that point to future work on massive PBHs. Firstly, our simulations are DM-only, which are suitable for studying large-scale structures, but they cannot accurately model the structure on the scale of individual halos. Moreover, without incorporating baryonic physics \citep[e.g., baryon-DM streaming motion, see][]{Kashlinsky2021PhRvL.126a1101K,Atrio2022,Boyuan2022MNRAS.514.2376L} in our simulations, it is challenging to model the growth of PBHs 
and their feedback effects on galaxies \citep[see, e.g.,][]{Pandey2018,Lu2021,Takhistov2022,DeLuca2023,Casanueva-Villarreal2024}. Additionally, with the current simulation volume, we do not produce any DM halos with masses $\gg 10^9\,\Msun$ at $z \sim 10$. To compare our model with JWST observations of high-redshift ($z \simeq 10$) galaxies \citep[e.g.,][]{Maisies:2022, Castellano2023ApJ...948L..14C}{}{}, we need to develop a mock galaxy catalogue over a much larger cosmic volume, with linear size of $L\gtrsim 10\ h^{-1} \rm Mpc$, capable of capturing the formation of more massive DM halos within the simulation box. With a larger simulation volume, we could also extend our study of the structure formation to lower redshifts ($z\simeq 1$), and use the NN statistics to test against sky surveys like SDSS-V \citep{SDSS2019BAAS...51g.274K} to better constrain the PBH abundance.

\begin{acknowledgments}
We would like to acknowledge the fruitful discussion with Nicola Bellomo, Steve Finkelstein and Florian K\"{u}hnel. 
BL is supported by the Royal Society University Research Fellowship. The authors acknowledge the Texas Advanced Computing Center (TACC) for providing HPC resources under allocation AST23026.
\end{acknowledgments}

%

\vspace{5mm}
\facilities{Lonestar6 (TACC)}


\software{astropy \citep{2013A&A...558A..33A,2018AJ....156..123A},  
          Colossus \citep{Diemer2018ApJCOLOSSUS}
          }



\newpage
\appendix
\section{Cosmic Web and Nearest Neighbor Statistics for other Runs}\label{AppendA}

\begin{figure}[htb!]
    \centering
    \includegraphics[width= \linewidth]{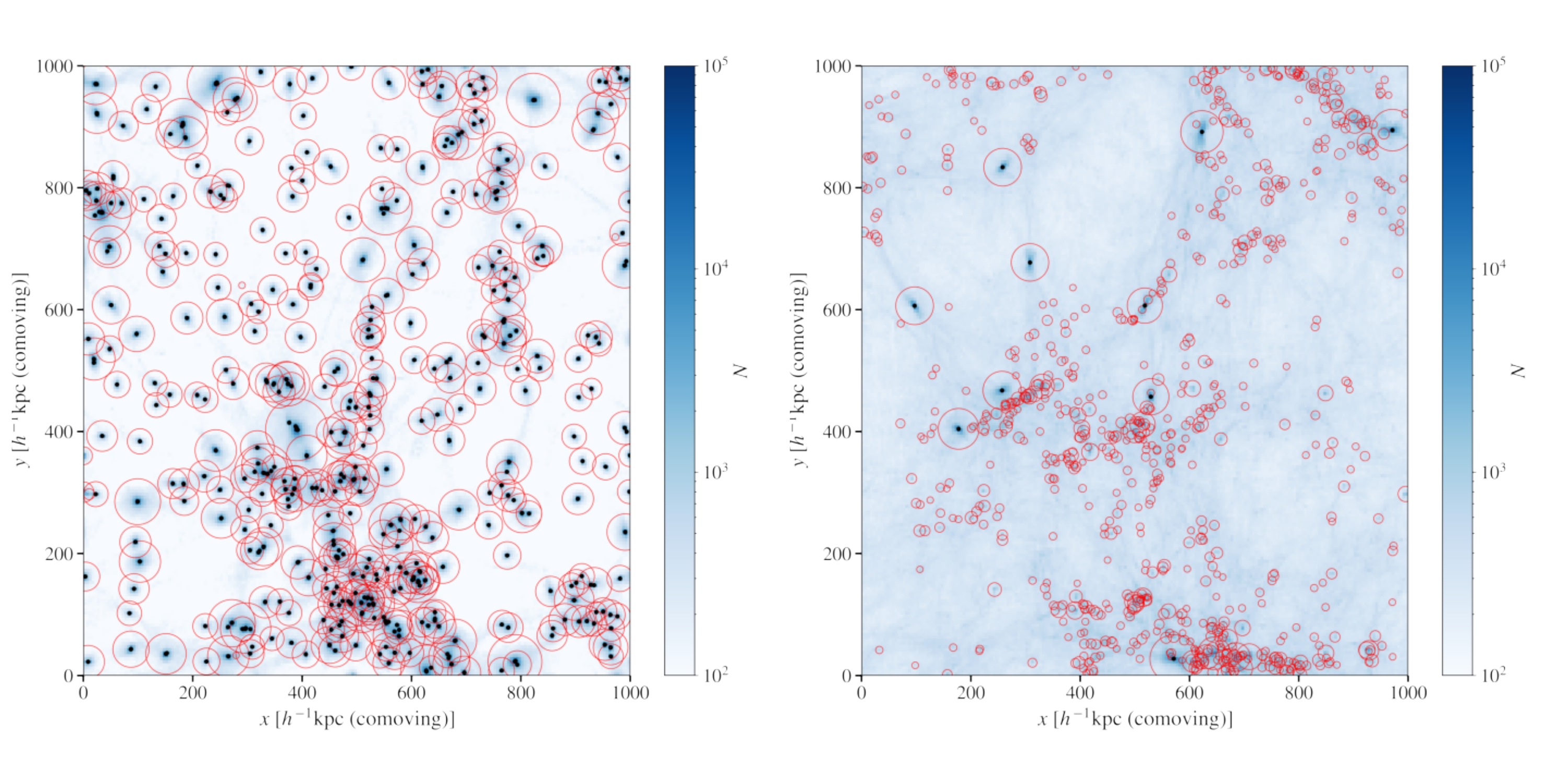}
    \vspace{-35pt}
    \caption{Structure formation for different PBH abundances. Similar to Fig.~\ref{fig:cswb}, we project our simulation box onto a 2D plane to depict the cosmic web at $z=10$ in comoving units. {\it Left panel:} Cosmic Web from the \texttt{PBH2} simulation run, with $f_{\rm PBH} = 10^{-2}$. {\it Right panel:} Cosmic Web from the \texttt{PBH4} simulation run, with $f_{\rm PBH} = 10^{-4}$. In both cases, all subhalos within their more massive hosts have been removed to better showcase the spatial distribution of halos and PBHs.}
    \label{fig:cswb_2}
\end{figure}

\begin{figure}[htb!]
    \hspace{-2cm}
    \includegraphics[width=1.22\textwidth]{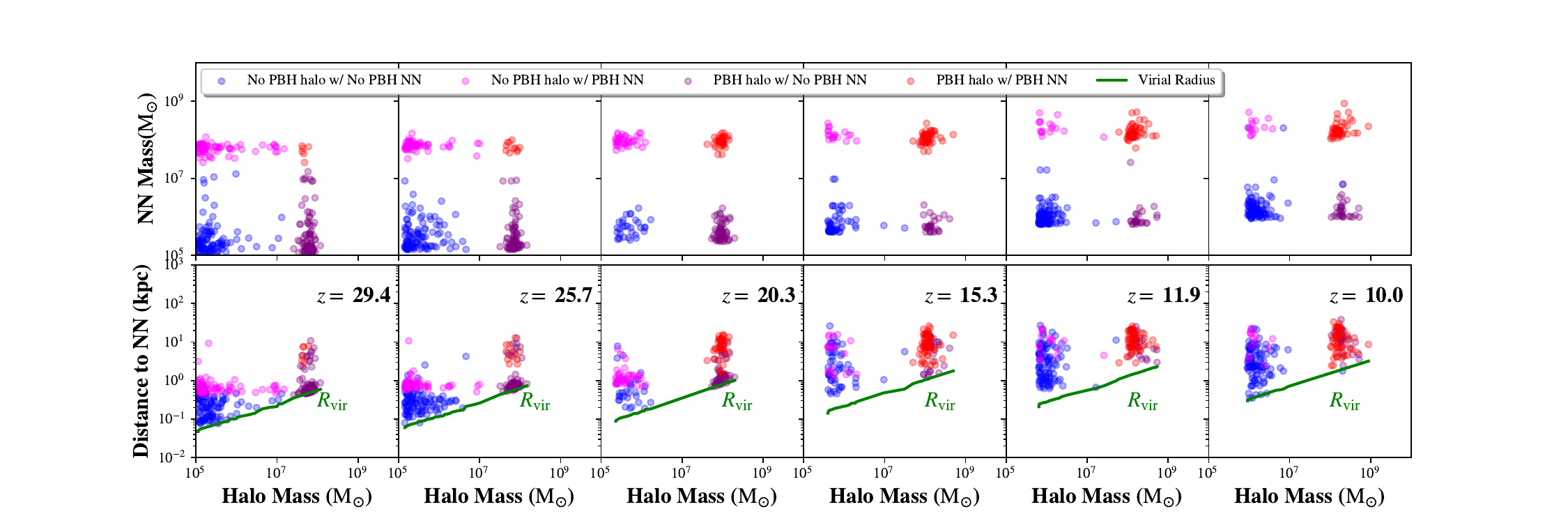}
    
    \vspace{-10pt}
    \caption{NN statistics for the \texttt{PBH\_PS} case, with a PBH mass fraction of $\fPBH = 10^{-3}$. Different from Fig.~\ref{fig:NNPBH3}, this simulation does not consider non-linear perturbations around PBHs in the initial conditions.}
    \label{fig:NNPBH_PS}
\end{figure}

\begin{figure}[htb!]
    \hspace{-2cm}
    \includegraphics[width=1.22\textwidth]{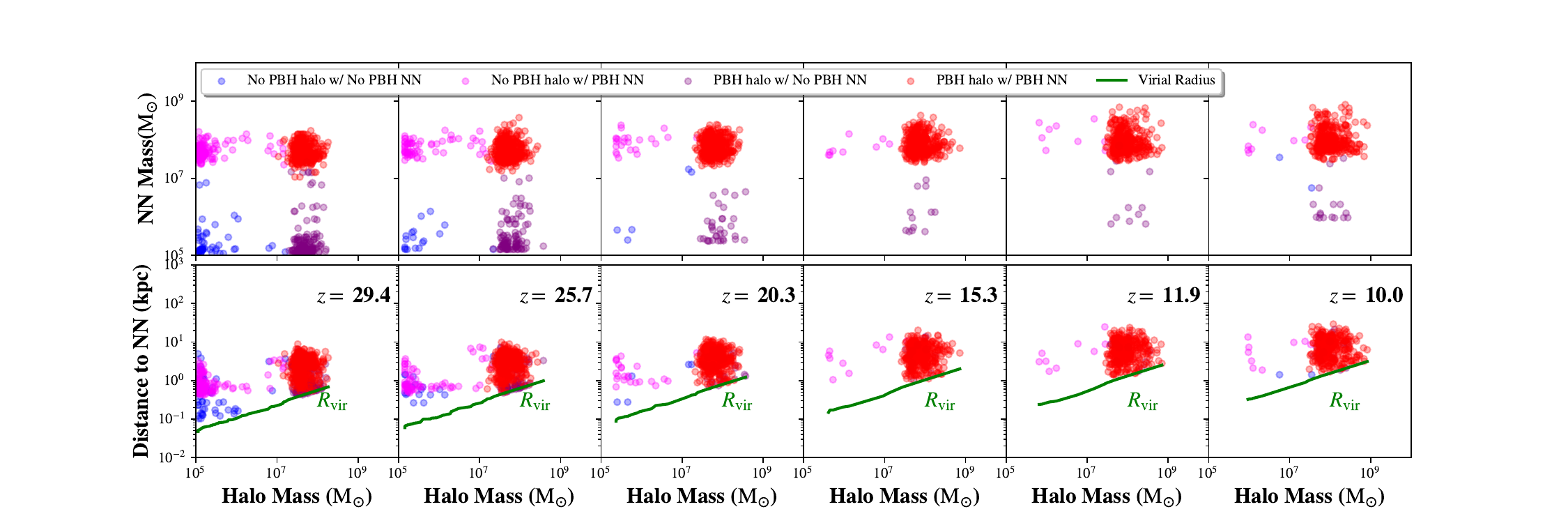}
    
    \vspace{-10pt}
    \caption{NN statistics for large PBH-abundance case. Similar to Fig.~\ref{fig:NNPBH_PS}, we analyze the \texttt{PBH2} run, with a PBH mass fraction of $\fPBH = 10^{-2}$. We employ the same convention for the  colors as before to differentiate halo groups.}
    \label{fig:NNPBH2}
\end{figure}

\begin{figure}[htb!]
    \hspace{-2cm}
    \includegraphics[width=1.22\textwidth]{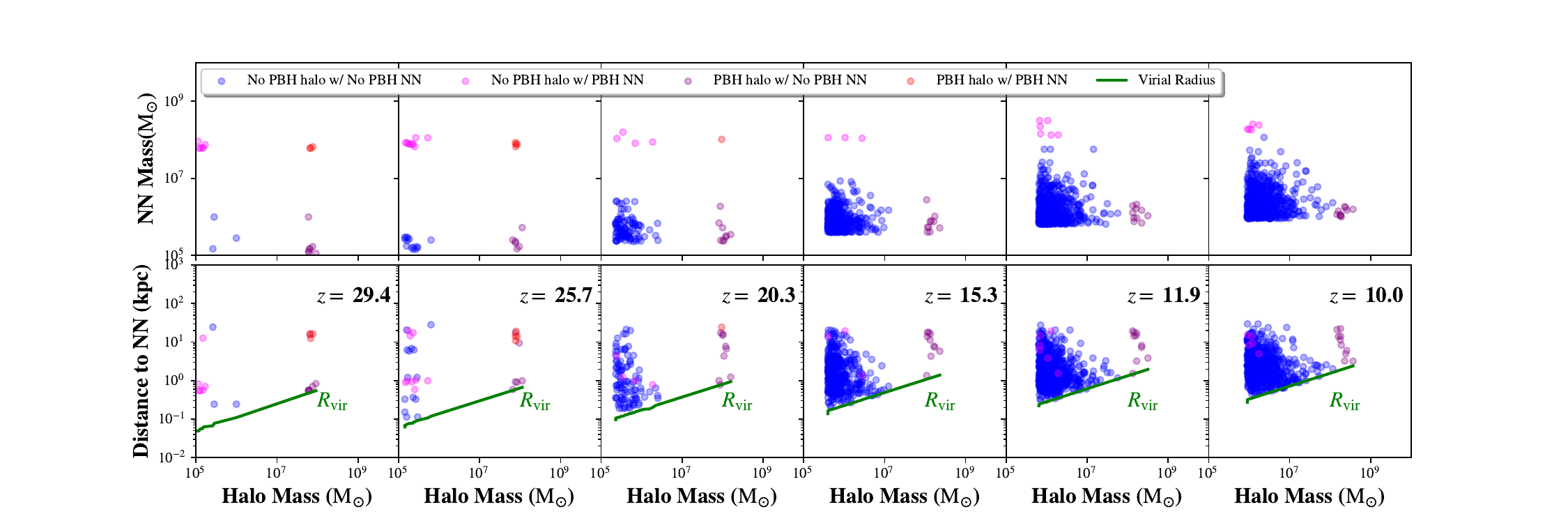}
    
    \vspace{-10pt}
    \caption{NN statistics for small PBH abundance. Similar to Fig.~\ref{fig:NNPBH_PS}, we plot the same quantities for the \texttt{PBH4} case, with a PBH mass fraction of $\fPBH = 10^{-4}$. We again use the same color convention to differentiate groups of halos. Note the convergence to the standard $\Lambda$CDM case, as expected for increasingly small PBH abundances. }
    \label{fig:NNPBH_4}
\end{figure}

\newpage

\bibliography{Main}{}
\bibliographystyle{aasjournal}



\end{document}